\documentclass[aps,prl,twocolumn,showpacs,nofootinbib,preprintnumbers]{revtex4}

\usepackage{color,amsmath,amssymb,subfigure}
\usepackage[dvips]{graphicx}

\begin{document}

\title{Measuring the extragalactic background light from very high energy
gamma-ray observations of blazars}

\author{Qiang Yuan$^1$}
\author{Hai-Liang Huang$^{1,2}$}
\author{Xiao-Jun Bi$^1$}
\author{Hong-Hao Zhang$^{2}$}

\affiliation{
$^{1}$Key Laboratory of Particle Astrophysics, Institute of High Energy
Physics, Chinese Academy of Sciences, Beijing 100049, P. R. China \\
}
\affiliation{$^{2}$School of Physics and Engineering, Sun Yat-Sen 
University, Guangzhou 510275, P. R. China}

\date{\today}

\begin{abstract}

The extragalactic background light (EBL) contains important information
about stellar and galaxy evolution. It leaves imprint on the very high 
energy $\gamma$-ray spectra from sources at cosmological distances
due to the process of pair production. In this work we propose to 
{\em measure} the EBL directly by extracting the collective attenuation 
effects in a number of $\gamma$-ray sources at different redshifts.
Using a Markov Chain Monte Carlo fitting method, the EBL intensities 
and the intrinsic spectral parameters of $\gamma$-ray sources are derived 
simultaneously. No prior shape of EBL is assumed in the fit. With this 
method, we can for the first time to derive the spectral shape of the 
EBL model-independently. Our result shows the expected features predicted 
by the present EBL models and thus support the understanding of the EBL origin.

\end{abstract}


\maketitle

The extragalactic background light (EBL) is the diffuse radiation from 
ultraviolet to far infrared wavelengths, spread isotropically in the 
universe (for a review of EBL, see \cite{2012arXiv1209.4661D,
2005PhR...409..361K}). The EBL originates from the radiative energy 
releases of all the stars, other extragalactic sources and diffuse 
emissions since the epoch of recombination. Therefore its intensity 
and spectral shape hold crucial information about the formation and 
evolution of stellar objects and galaxies throughout the cosmic history.
The EBL is one of the fundamental quantities in cosmology.

Direct measurement of EBL is, however, very difficult due to the 
contamination of the foreground emission from the solar system
zodiacal light and the Galactic stellar and interstellar emissions
\cite{2001ARA&A..39..249H}. Technically, it also requires the absolute 
calibration of the instruments, and the understanding all measurement
uncertainties. Given the difficulties, direct measurements provide 
just lower and upper limits of EBL intensity. A strict lower limit 
on the EBL intensity is provided by the integrated light from resolved 
galaxies, e.g. in optical by the Hubble Space Telescope 
\cite{2000MNRAS.312L...9M} and in infrared by the Spitzer telescope 
\cite{2004ApJS..154...39F}. The upper limit can be derived from the 
absolute measurement of EBL within its errors \cite{2012arXiv1209.4661D}. 
The allowed range is shown by the shadow region in the following figures.

Extreme efforts had been paid to calculate the EBL intensity
\cite{2005AIPC..745...23P,2006ApJ...648..774S,2008A&A...487..837F,
2010A&A...515A..19K,2011MNRAS.410.2556D,2010ApJ...712..238F,
2012MNRAS.422.3189G}. The models generally include two distinctive 
processes, that the UV and optical component of EBL is the integral of 
starlight over all epochs and the infrared component is due to the 
process of absorption and re-emission of starlight by the interstellar 
dust. These models agree on the overall EBL shape, including two maxima 
at $\sim 1 \mu$m by starlight and at $\sim 100 \mu$m by dust.
However, since the detailed EBL model depends on many factors,
such as the star formation history, the stellar initial mass function, 
the evolution of metallicity, the energy released by AGN, the size 
distribution and composition of dust grains, different models keep 
a large diversity.

There is another indirect but effective way to study the EBL by 
observation of very high energy (VHE) $\gamma$-rays. The VHE 
$\gamma$-rays from extragalactic sources are attenuated by the 
process of electron/positron pair production, $\gamma_{\rm VHE}+
\gamma_{\rm EBL}\to e^+e^-$, when propagating to the Earth
(e.g., \cite{Nikishov1962,1966PhRvL..16..252G,1992ApJ...390L..49S}).
With the rapid development of ground based $\gamma$-ray imagining 
atmospheric Cerenkov telescopes (IACT), quite a few VHE $\gamma$-ray 
sources from cosmological distances have been detected, most of which 
are blazars, a subgroup of active galactic nuclei (AGN), with 
relativistic jet pointing towards the observer. With assumption of the 
intrinsic blazar spectra we can set an upper limit of the EBL intensity 
by comparing the observed spectra with the intrinsic spectra
\cite{1992ApJ...390L..49S}. The observations of blazars H 2356-309 and 
1ES 1101-232 at redshifts $z=0.165$ and $z=0.186$ respectively by HESS 
has set a strong upper limit of EBL, close to the lower limit set by 
galaxy counts, at the near infrared wavelength \cite{2006Natur.440.1018A}.
The MAGIC observation of 3C 279 at $z=0.536$ set upper limit at the 
optical band \cite{2008Sci...320.1752M}. In \cite{2007A&A...471..439M} 
Mazin and Raue gave a comprehensive study of EBL based on eleven blazars 
over a redshift range from $0.03 - 0.18$. They explored a large number 
of hypothetical EBL scenarios and set robust constraints on EBL over a 
wide wave-length range. With the Fermi observation of blazar spectra at 
GeV to $\sim 100$ GeV more stringent constraints on EBL are recently given 
by \cite{2009ApJ...698.1761F,2010ApJ...723.1082A,2010ApJ...714L.157G,
2010ApJ...715L..16M,2011ApJ...733...77O,2012A&A...542A..59M}.
Those studies seem to indicate that the Universe is more transparent 
than we had expected.

The power of this method to study EBL is limited due to the fact that 
the intrinsic spectrum of each blazar is unknown. Therefore it is hard 
to disentangle the absorption effect by EBL from the intrinsic emission 
nature for a specific observation. The usual practice in the literature 
is to reconstruct the blazar intrinsic spectrum from the observation
by first assuming an EBL model. The EBL model is rejected if it results
in an unphysical intrinsic spectrum, for example, the reconstructed 
intrinsic spectrum follows a power law with an extremely hard spectral 
slope or even shows an exponential rise at the high energy end. 
Recently with large sample of $\gamma$-ray blazars, the EBL intensities 
were derived through a likelihood fit with given spectral template of
the EBL \cite{2012Sci...338.1190A,2012arXiv1212.3409H}.

Considering the quickly accumulating number of $\gamma$-ray sources 
observed by IACTs and Fermi, which will be improved essentially by the 
future Cerenkov telescope array (CTA) \cite{2011ExA....32..193A,
2012arXiv1201.3276H,2012arXiv1205.1459Z}, we propose to {\em measure} 
the EBL directly by extracting the collective absorption effects in a 
number of $\gamma$-ray sources at different redshifts.
In this work we demonstrate the capability of this method by adopting
the Markov Chain Monte Carlo (MCMC) fitting to known data to extract
the parameters of the intrinsic spectra of a series of TeV blazars, as
well as the EBL intensities simultaneously. Different from the previous 
studies in the literature, we make no assumption of the EBL model in our 
fitting. Instead the EBL is divided into many discrete energy bins and 
the energy density in each bin is fitted as a free parameter $\xi_i$.

The observed VHE $\gamma$-ray spectrum after absorption by the EBL is
commonly expressed as
\begin{equation}
F_{\rm obs}(E) = e^{-\tau(E,z)} F_{\rm int}(E)\ ,
\label{obs}
\end{equation}
where $F_{\rm int}(E)$ is the intrinsic spectrum of the source at redshift 
$z$. The strength of the attenuation by EBL is described by the optical 
depth $\tau(E,z)$ as a function of energy $E$ and the source redshift $z$.
The optical depth $\tau$ is expressed as \citep{1967PhRv..155.1404G}
\begin{eqnarray}
\tau(E,z)&=&\int_0^z {\rm d}l(z')\int_{-1}^{+1} {\rm d}\mu \frac{1-\mu}{2} 
\nonumber\\
&&\cdot\int_{\epsilon'_{\rm thr}}^\infty {\rm d}\epsilon' n'(\epsilon',z')
\sigma(E',\epsilon',\mu),
\label{tau}
\end{eqnarray}
where variables with prime are the quantities at redshift $z'$,
${\rm d}l=c{\rm d}t=\frac{c}{H_0}\frac{{\rm d}z'}{(1+z')
\sqrt{\Omega_M(1+z')^3+\Omega_{\Lambda}}}$ is the differential path
traveled by the VHE photon, $\mu=\cos\theta$ with $\theta$ the angle
between the momenta of VHE and EBL photons, $n'(\epsilon',z') = 
n(\epsilon'/(1+z'),z=0)(1+z')^3 $ is the EBL number density at 
redshift $z'$, and $\sigma$ is the pair production cross section. 
$\epsilon'_{\rm thr}$ is the threshold energy for $\gamma$-ray energy 
$E'=E(1+z')$ with an angle $\cos\theta=\mu$ with the EBL photon. The 
cross section is peaked at a wavelength $\lambda/\mu {\rm m} \sim 
1.24 E/{\rm TeV}$ \cite{2000A&A...359..419G}. Therefore the observation 
of VHE $\gamma$-ray spectra can probe EBL at the wavelength from optical 
to far infrared, while it is not sensitive to UV band. The cosmological 
parameters used in this work are $\Omega_M=0.274$, $\Omega_{\Lambda}=
1-\Omega_M$, $H_0=70.5$ km s$^{-1}$ Mpc$^{-1}$ \cite{2009ApJS..180..330K}.

In the fitting the intrinsic spectra of blazars $F_{\rm int}$ are 
parameterized by a power-law ($F\propto E^{-\alpha}$) or log-parabolic
($F\propto E^{-\alpha-\beta\log E}$) function, with two or three free 
parameters for each source. The blazar spectrum is usually explained 
by the synchrotron-self-Compton (SSC) model, which shows a concave 
$\gamma$-ray spectrum. Such a spectrum can be formulated by the 
log-parabolic function. If the measured energy range is not too wide
the simple power-law can give a quite good description. In the 
following we will compare the results with the two spectral forms.

No prior assumption about EBL shape is adopted in this study.
The EBL intensities are divided into $10$ bins logarithmically between
$0.1$ and $100$ $\mu$m. Within each bin the intensity $\nu I_{\nu}$
is assumed to be a constant $\xi_i$. Then we can fit the 10 $\xi_i$s,
according to Eq. (\ref{obs}), from a set of observed $\gamma$-ray 
spectra $F_{\rm obs}(E)$.

We have adopted seven blazars in this study, which have relatively precise 
spectral measurements. The sources adopted are listed in Table 
\ref{table:sample}. Since the redshifts of these sources are less than 
$0.2$, we neglect the evolution of the EBL \cite{2006Natur.440.1018A}.

\begin{table}[!htb]
\centering
\caption{Source sample information}
\begin{tabular}{cccc}
\hline \hline
 Name & redshift & Experiment & Reference \\
\hline
 Mkn 421      & 0.031 & VERITAS & [33] \\
 Mkn 501      & 0.034 & HEGRA   & [34] \\
 1ES 1959+650 & 0.047 & HEGRA   & [35] \\
 PKS 2005-489 & 0.071 & HESS    & [36] \\
 PKS 2155-304 & 0.116 & HESS    & [37] \\
 H 2356-309   & 0.165 & HESS    & [38] \\
 1ES 1101-232 & 0.186 & HESS    & [16] \\
\hline
\end{tabular}
\label{table:sample}
\end{table}

\begin{figure}[!htb]
\centering
\includegraphics[width=0.45\textwidth]{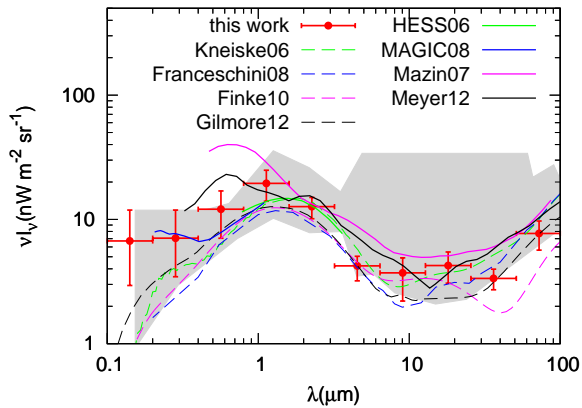}
\caption{The best-fitting results of the EBL intensities in 10 wavelength 
bins. The intrinsic spectra of blazars are assumed to be log-parabolic
and the low state of Mkn 421 spectrum by VERITAS is adopted. The solid
lines show the upper limits derived by the $\gamma$-ray observations of 
blazars, HESS06 \cite{2006Natur.440.1018A}, Mazin07
\cite{2007A&A...471..439M}, MAGIC08 \cite{2008Sci...320.1752M},
and Meyer12 \cite{2012A&A...542A..59M}, and the dashed lines show the 
model predictions of Kneiske06 \cite{2004A&A...413..807K}, Franceschini08 
\cite{2008A&A...487..837F}, Finke10 \cite{2010ApJ...712..238F} and 
Gilmore12 \cite{2012MNRAS.422.3189G}.
}
\label{fig1}
\end{figure}

We employ the MCMC algorithm to do this global fit, which is very 
efficient for the minimization in high-dimensional parameter space 
\cite{2002PhRvD..66j3511L}. Physical constraints on the parameters 
are adopted. In models of diffusive shock acceleration 
of electrons in the blazar jets the deduced $\gamma$-ray spectra are 
strongly constrained with power law index $\alpha$ larger than 
$1.5$ \cite{2001RPPh...64..429M}. The parameter $\beta$ is restricted
in $[0,1]$, and we shall test in the following that relaxing the range
of $\beta$ does not change the results significantly. The EBL intensities 
$\xi_i$ are restricted in the lower and upper limits set by direct 
measurements \cite{2012arXiv1209.4661D}. 

Fig. \ref{fig1} shows the fitting result assuming log-parabolic intrinsic 
spectra of balzars. There are seven states of Mkn 421 observed by VERITAS
\cite{2011ApJ...738...25A}, and the low state spectrum (sample 2 as defined
below) is adopted in this fit due to its wide energy coverage. 
For comparison, several recent model predictions of the EBL at $z=0$ 
\cite{2004A&A...413..807K,2008A&A...487..837F,2010ApJ...712..238F,
2012MNRAS.422.3189G} are plotted by the dashed lines in the figure. 
Our fitting result agrees very well with the model predictions for 
$\lambda>1$ $\mu$m. The expected peaks at $\sim 1\mu$m and $\sim 100\mu$m 
are also shown in this model-independent fit. It is very interesting 
that the indirect {\em measurement} of EBL provides strong support to 
the present understanding of EBL origin. The solid lines in the same plot 
show several current upper bounds on the EBL intensities according to the 
$\gamma$-ray observations of blazars \cite{2006Natur.440.1018A,
2007A&A...471..439M,2008Sci...320.1752M,2012A&A...542A..59M}. 

\begin{figure*}[!htb]
\centering
\includegraphics[width=0.45\textwidth]{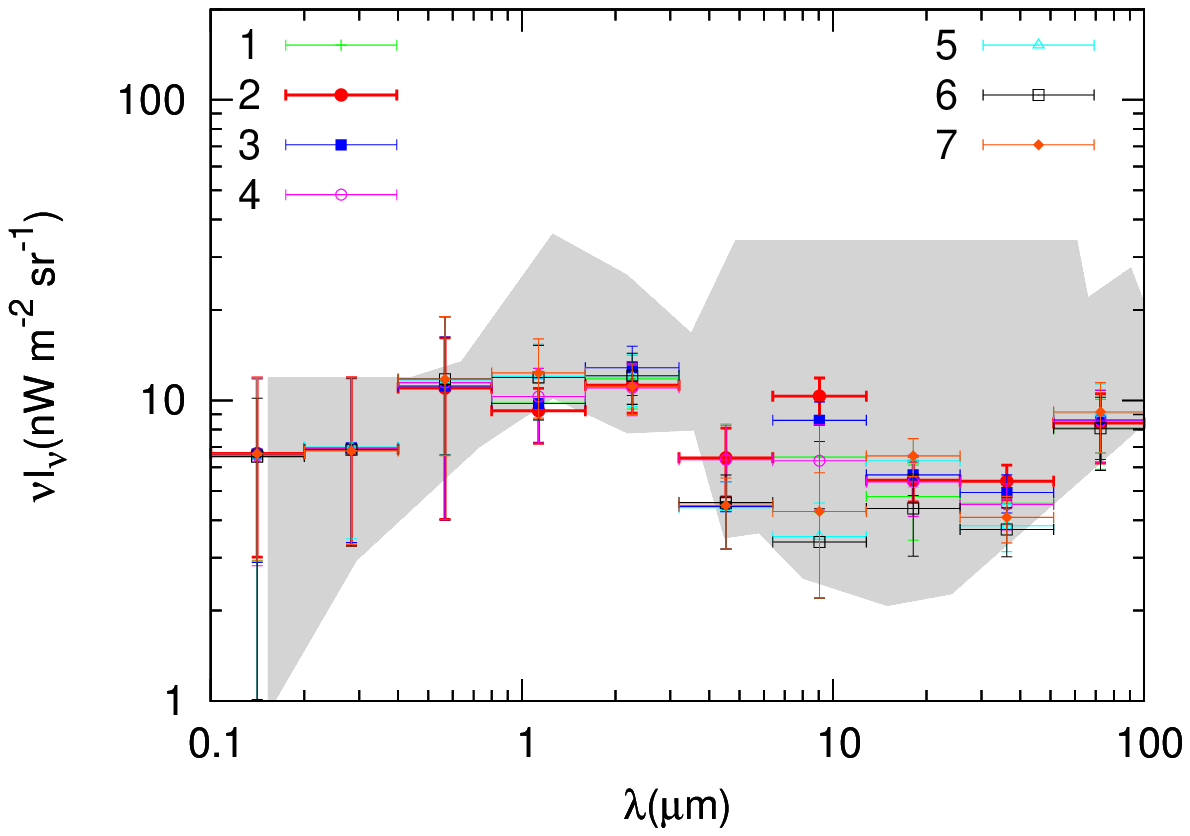}
\includegraphics[width=0.45\textwidth]{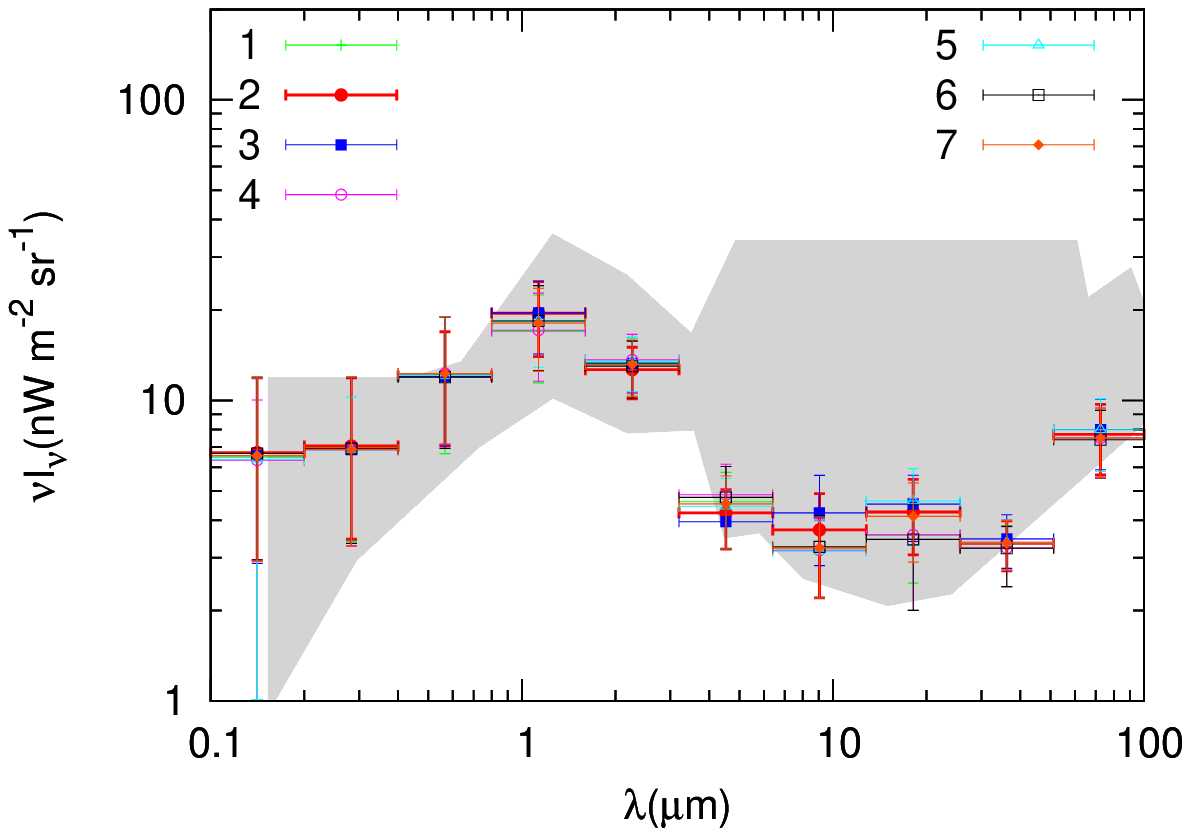}
\caption{The best-fitting results of the EBL intensities in 10
wavelength bins, for the assumptions of power-law (left) and
log-parabolic (right) intrinsic spectra of the sources. Different
symbols are for different data sets (see the text for details).
} \label{test}
\end{figure*}

\begin{table*}[!htb]
\centering
\caption{Best-fit $\chi^2$/d.o.f. values}
\begin{tabular}{cccccccccc}
\hline \hline
 & 1 & 2 & 3 & 4 & 5 & 6 & 7 \\
\hline
  power-law    & $90.8/55$ & $89.4/58$ & $73.6/59$ & $87.6/56$ & $64.7/50$ & $46.8/50$ & $63.4/53$ \\
  log-parabolic& $54.7/48$ & $60.2/51$ & $61.8/52$ & $53.1/49$ & $61.2/43$ & $46.1/43$ & $52.9/46$\\
  \hline
\end{tabular}
\label{table:chi2}
\end{table*}

In the following we discuss the robustness of the method. The largest 
uncertainty comes from the unknown nature of the source emission. 
Fortunately Mkn 421 provides a perfect template to test the method. 
The observations by VERITAS of Mkn 421 covered seven states from very 
low to very high states from 2006 to 2008 \cite{2011ApJ...738...25A}. 
We can test the convergence of EBL by the fitting procedure using the 
different states data at the same source. Fig. \ref{test} shows the 
results by combining the other 6 sources with the different states of 
Mkn 421 (referred as sample 1-7 from very low to very high state), 
for assumption of power-law (left) and log-parabolic (right) intrinsic 
spectra of the sources. For power-law intrinsic spectra, the results 
show moderate diversity among different data set, while for the 
log-parabolic intrinsic spectra the results converge quite well.

The reduced $\chi^2_r=\chi^2$/d.o.f. are listed in Table
\ref{table:chi2}. It is shown that for power-law source spectra
the reduced $\chi^2_r$ varies for different data samples. In most
of these cases the $\chi^2$ values are too large. For the
log-parabolic source spectra, the reduced $\chi^2_r\sim1$ for
almost all of the data samples. This means the SSC model predicted
log-parabolic function can give a quite well description to the
intrinsic $\gamma$-ray spectra. For Mkn 421, some of the seven
observations span the energy range for more than $1.5$ orders of
magnitude \cite{2011ApJ...738...25A}, thus the simple power-law
is not a good description.

\begin{figure}[!htb]
\centering
\includegraphics[width=0.45\textwidth]{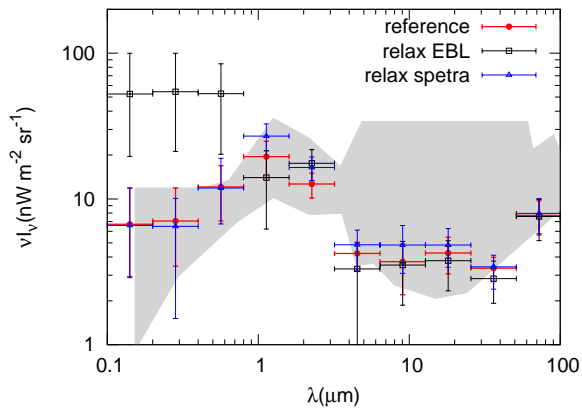}
\caption{Comparison of the results for relaxing fitting ranges of
the EBL intensities or the spectral parameters. The reference case
is the log-parabolic fit to data sample 2. } \label{relax}
\end{figure}

Then we relax our constraints on the parameter space. The EBL
intensities are relaxed to be within $[1,100]$ nW m$^{-2}$ sr$^{-1}$. 
The spectral parameters are relaxed as $\alpha\geq2/3,\,0\leq\beta\leq2$ 
respectively. The results are shown in Fig. \ref{relax}. We notice 
that the current adopted data sample can not constrain the EBL with 
$\lambda<1$ $\mu$m. The UV-optical band can be constrained by sources 
at high redshift and lower energy, such as the 3C 279 at $z=0.536$ 
\cite{2008Sci...320.1752M}. However, since only 5 points of 3C 279 
was given by MAGIC it has no help to improve the fit because to include 
it we need to introduce 3 additional source parameters.

In summary we propose to {\em measure} the EBL from the VHE $\gamma$-ray
data by a global fitting method. Both the intrinsic spectral parameters 
and the EBL intensities are fitted simultaneously using an MCMC algorithm,
without any assumption of the spectral shape of EBL. With a log-parabolic
VHE $\gamma$-ray spectra the fitting shows well convergence for the
EBL intensities. The EBL intensities are close to the lower bound of EBL 
set by galaxy counts and are consistent with the recent EBL models. 
With the greatly improved number of VHE $\gamma$-ray sources by the 
future CTA the EBL can be determined with much higher precision.

\acknowledgments 

This work is supported by the Natural Sciences Foundation of China under
grant Nos. 11075169, 11105155, 11135009 and 11005163.

\bibliography{refs}
\bibliographystyle{apsrev}

\end{document}